\documentclass[english,aps,prd,11pt,superscriptaddress,preprintnumbers,nofootinbib]{revtex4}
\usepackage[T1]{fontenc}
\usepackage[latin9]{inputenc}
\usepackage{verbatim}
\usepackage{amsmath}
\usepackage{dcolumn}
\usepackage{graphicx}

\allowdisplaybreaks

\newcommand{\be}{\begin{equation}}
\newcommand{\ee}{\end{equation}}
\newcommand{\bea}{\begin{eqnarray}}
\newcommand{\eea}{\end{eqnarray}}
\newcommand{\ep}{\epsilon}

\usepackage{babel}
\begin{document}
\global\long\def\order#1{\mathcal{O}\left(#1\right)}
\global\long\def\d{\mathrm{d}}
\global\long\def\P{P}
\global\long\def\amp{{\mathcal M}}
\preprint{}

\title{
\vskip-3cm{\baselineskip14pt
    \begin{flushleft}
      \normalsize TTP14-026
  \end{flushleft}}
  \vskip1.5cm
Two-loop helicity amplitudes  for the production of two off-shell 
electroweak bosons   in quark-antiquark collisions
}

\author{Fabrizio Caola}
\email{fabrizio.caola@kit.edu}
\affiliation{ Institute for Theoretical Particle Physics, KIT, Karlsruhe, Germany}

\author{Johannes M. Henn}
\email{jmhenn@ias.edu}
\affiliation{Institute for Advanced Study, Princeton, NJ 08540, USA, USA}
\author{Kirill Melnikov\footnote{On leave of absence from
Department of Physics and Astronomy, Johns Hopkins University, Baltimore,
MD, USA}
}
\email{melnikov@pha.jhu.edu}
\affiliation{ Institute for Theoretical Particle Physics, KIT,
Karlsruhe, Germany}
%Department of Physics and Astronomy, Johns Hopkins University, Baltimore, USA
%}

\author{Alexander V. Smirnov}
\email{asmirnov80@gmail.com}
\affiliation{Scientific Research Computing Center, Moscow State University, 119991 Moscow, Russia}
\author{Vladimir A. Smirnov}
\email{smirnov@theory.sinp.msu.ru}
\affiliation{
Skobeltsyn Institute of Nuclear Physics of Moscow State University, 119991 Moscow, Russia
}

\begin{abstract}
Knowledge of  two-loop QCD amplitudes for  processes 
$q \bar q' \to V_1 V_2  \to (l_1 \bar l'_{1})  (l_2 \bar l'_2)$
is important for improving  the theoretical description of
four-lepton production  in hadron collisions.  In this paper we compute these helicity  amplitudes 
for all intermediate vector bosons,   $V_1 V_2 = \gamma^* \gamma^*$, 
$W^+W^-$, $ZZ$, $W^\pm Z$, $W^\pm \gamma^*$, including off-shell effects and decays to leptons. 

\end{abstract}

\maketitle

\thispagestyle{empty}

\newpage

\section{Introduction} 
Production of vector boson pairs   at the LHC is an interesting  process for a variety of 
reasons. For definiteness, let us  consider the $W^+W^-$ final state. 
In that case,  ATLAS and CMS collaborations have recently observed  that the $W^+W^-$ production 
cross section  is about twenty percent higher than existing  theoretical predictions \cite{atlas7,cms7,cms8}.   
This observation prompted  speculations about the possibility 
to explain this excess  by physics beyond the Standard Model \cite{bsm} and, at the same time,  strongly 
emphasized the need to improve predictions for $W^+W^-$ production 
  within the Standard Model itself~\cite{Meade:2014fca, Jaiswal:2014yba}.  
In addition, production of $W^+W^-$ pairs 
is an important process for studying anomalous couplings of electroweak gauge boson. Although current limits 
are already quite impressive \cite{limits}, it is clear that  
studies of  anomalous gauge boson couplings 
will intensify  once  the  LHC Run II is underway.  Making use of higher experimental precision will 
require improved modeling of $W^+W^-$ production in the Standard Model.  
Finally, the $pp \to W W^*$ process   with one $W$-boson on the mass shell and 
the other off the mass-shell, is an important background to  Higgs boson production in 
$pp \to H \to WW$ channel. Better understanding of this background should allow improved measurements 
of Higgs boson couplings to $W$-bosons -- including the anomalous ones --  in the next run of the LHC.  
Similar arguments can be given for processes  with other vector bosons in the final state.

It follows from these examples, that 
higher theoretical accuracy for vector-boson pair production 
in hadron collisions is essential. It can be achieved by  extending  
existing computations of cross sections and kinematic distributions of $pp \to V_1 V_2$ processes 
\cite{Jaiswal:2014yba,Meade:2014fca,
nloqcd,shower,
resum,ax,gglo}
to 
next-to-next-to-leading order (NNLO) in perturbative QCD.\footnote{First NNLO QCD results for electroweak boson pair production 
have recently appeared, see e.g. Refs.\cite{Cascioli:2014yka,babis,tg1}.}
In particular, predictions 
for   fiducial volume cross sections, where kinematic restrictions on final state 
particles are taken into account exactly, are crucial.   
A NNLO QCD prediction for $pp \to V_1 V_2 $  needs three ingredients: {\it i}) 
real-emission matrix elements  for $ q \bar q' \to V_1 V_2 gg$, $qg \to V_1 V_2  qg$ and  $q \bar q \to V_1 V_2  q\bar q$;
{\it ii}) one-loop matrix elements for $ q \bar q \to V_1 V_2 g$ and $ q g  \to V_1 V_2  q$ 
and, finally, {\it iii}) two-loop amplitudes for tree-level process $q \bar q' \to V_1 V_2 $.  
Once these three ingredients become available, they need to be put together in a self-consistent manner using 
existing methods for fully differential NNLO computations \cite{nnlo_tech}.
For the $V_1 V_2 $ production, the major unknown 
is the two-loop amplitude for $q \bar q'  \to V_1 V_2 $; the goal of this paper is to provide it.

The remainder of the paper is organized as follows. In Section~\ref{section1} we explain the general setup of the calculation. 
Since two-loop scalar  master integrals required for this computation have been recently computed 
for equal electroweak boson masses in Refs.~\cite{Gehrmann:2014bfa,Gehrmann:2013cxs} and for unequal masses in 
Refs.~\cite{planar,nonplanar}, we mainly focus on the procedure 
that allows us to express the various contributions to  scattering amplitudes in terms 
of these integrals. In Section~\ref{results}  we describe  checks on our computation, evaluate  
the scattering amplitudes numerically and discuss  numerical stability of the results.
We conclude in Section~\ref{concl}.

\section{The setup of the computation} 
\label{section1}

We are interested in the production of a four-lepton final states in proton collisions
$p p \to (l_1 \bar l'_{1})  (l_2 \bar l'_2)$. The four-lepton final states can be 
produced in three  distinct ways:
\begin{itemize} 
\item   through the production of a {\it pair}  of off-shell vector bosons,  $ p p  \to V_1 V_2  \to (l_1 \bar l'_{1})  (l_2 \bar l'_2)$;
\item   through the production 
of a {\it single}  vector boson that  decays to {\it two} vector bosons that, in turn,  decay to lepton pairs, 
 $ p p  \to V_3 \to V_1 V_2  \to (l_1 \bar l'_{1})  (l_2 \bar l'_2)$;
\item   through the production of a {\it single}  vector boson that decays into a pair of leptons (one of them of-shell), 
followed by  the emission of a vector boson by an of-shell  lepton, e.g.  
$pp \to V_3 \to l_3^*  \bar l'_2 \to  ( l_3^* \to V_1 l_2 ) \bar l'_2 \to  (l_1 \bar l'_{1})  (l_2 \bar l'_2)$.  
\end{itemize}

Each of these three processes depends on different combinations of electroweak couplings;  therefore, 
they are separately invariant under QCD gauge transformations and we can compute QCD corrections 
to each of them separately. We note in this respect  that QCD corrections to processes 
mediated by single gauge-boson production are simple since they are directly related to QCD 
corrections  to the quark form factor of the vector current 
$F_V(s)$ whose perturbative expansion through  NNLO is well-known \cite{ffnnlo}.

However, computation of  NNLO QCD corrections to  process 
$ p p  \to V_1 V_2  \to (l_1 \bar l'_{1})  (l_2 \bar l'_2)$, where vector bosons couple only to fermions 
and not to other vector bosons or primary leptons,  
is non-trivial.  Calculation of this contribution to two-loop QCD  amplitudes for four-lepton production 
processes is the main focus of this paper.   

We  consider quark-antiquark annihilation in  to vector bosons
\be
\label{eq1} 
q(p_1 ) \bar q'(p_2) \to V_1(p_3) V_2(p_4) \to  (V_1(p_3) \to l (p_5) \bar l'(p_6) )\;   (V_2(p_4) \to l ( p_7)  \bar l'(p_8) ),
\ee
and work in the  approximation where quarks of the first two generations are  massless and 
quarks of the third generation are consistently neglected.  We also set the CKM matrix to an identity 
matrix.   Since we work with massless quarks, helicity is a conserved  quantum number; therefore, once 
the helicity of the incoming quark is specified, the helicity of the incoming anti-quark is 
completely fixed. We will use this observation when writing the amplitude for quark-antiquark 
annihilation process in Eq.(\ref{eq1}). 

\begin{figure}[t]
  \centering
  \includegraphics[angle=0,width=0.8\textwidth]{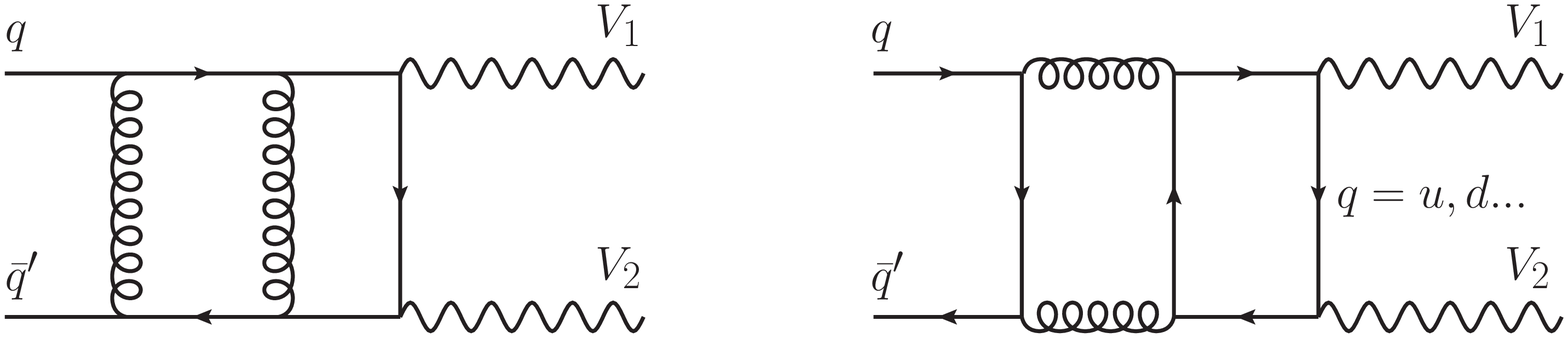}\\
  \caption{Representative diagrams that contribute to two-loop amplitude 
for  vector boson  production in hadron collisions. }
  \label{fig1}
\end{figure}

The partonic process in Eq.(\ref{eq1}) can proceed in two different ways since the vector bosons can either 
couple directly to external fermions $q \bar q'$ or to closed loops of  virtual fermions, see  Fig.\ref{fig1}. 
Consequently, 
we write the scattering amplitude for process in Eq.(\ref{eq1}) as 
\be
\begin{split} 
& {\cal M}(\lambda_q,\lambda_5,\lambda_7) = i\left ( \frac{g_W}{\sqrt{2}} \right )^4 
  \delta_{i_1 i_2} {\cal D}_3(p_3){\cal D}_4(p_4)  C_{l,V_2}^{\lambda_7} C_{l,V_1}^{\lambda_5} 
\epsilon_3^\mu(\lambda_5)  \epsilon_4^\nu(\lambda_7) 
\\
& \times \left [ 
C_{{\bar q}',V_2}^{\lambda_q} C_{q,V_1}^{\lambda_q}   {\cal A}^{(d)}_{\mu \nu}(p_1^{\lambda_q},p_3, p_4, p^{-\lambda_q}_2) 
+
C_{ {\bar q}',V_1}^{\lambda_q} C_{q,V_2}^{\lambda_q}   {\cal A}^{(d)}_{\nu \mu}(p_1^{\lambda_q},p_4, p_3, p^{-\lambda_q}_2) 
\right. 
\\
& \left. 
+   C_{V_1 V_2}^{n_g} {\cal A}^{n_g}_{\mu \nu}(p_1^{\lambda_q},p^{-\lambda_q}_2; p_3, p_4) 
\right ], 
\label{eq2}
\end{split}
\ee
where $g_W = e/\sin \theta_W$ is the $SU(2)$ weak coupling,  
${\cal D}_i = 1/(p_i^2 - m_{V_i}^2 + i m_{V_i} \Gamma_{V_i})$ is the $V_i$-boson propagator,
$\lambda_q, \lambda_5, \lambda_7$ are helicities  of the incoming quark and outgoing leptons, 
respectively,  
$C_{{\bar q}',V_2}^{\lambda_q}, C_{q,V_1}^{\lambda_q}$ 
and $C_{l,V_2}^{\lambda_7} C_{l,V_1}^{\lambda_5} $ are 
helicity-dependent 
couplings of vector bosons to quarks and leptons,
 and  $\epsilon_{3,4}$  are matrix elements for leptonic decays of $V_1$ and $V_2$ that we will 
specify shortly.  The amplitudes ${\cal A}^{(d)}$ describe direct coupling of the vector bosons 
to external fermions and the amplitude ${\cal A}^{n_g}$ describes  contributions of diagrams where 
vector bosons  couple 
to loops of virtual fermions, see Fig.~\ref{fig1}.  The factor $C_{V_1 V_2}^{n_g}$ involves sums over couplings of virtual fermions 
to gauge bosons. 

We note that  full amplitude for $q \bar q'  \to V_1 V_2$ in Eq.(\ref{eq2}) 
is written as the sum of {\it two} ${\cal A}^{(d)}$ amplitudes where  two vector bosons appear in different order.  These amplitudes
are, therefore,  quantities similar to ordered or primitive amplitudes often used in perturbative QCD computations. 
For us they are useful because we only need to compute one of them and then 
obtain the other one from $p_3 \leftrightarrow p_4$ permutation.  We stress that amplitudes 
${\cal A}^{(d)}$ and ${\cal A}^{(n_g)}$ 
in Eq.(\ref{eq2})  are computed assuming that vector bosons couple to a single quark generation 
and that  the $Vqq$  coupling is vector-like with  unit coefficient, i.e. $\gamma_\mu$.  
As we explain below, such amplitudes  are sufficient to obtain physical 
amplitudes for pair production of all electroweak gauge  bosons. 

We begin by discussing the parametrization of ${\cal A}^{(d)}_{\mu \nu} \epsilon_3^\mu \epsilon_4^\nu$ 
amplitude in terms of Lorentz-invariant form factors. 
A representative diagram that contributes to ${\cal A}^{(d)}$ at two loops in perturbative QCD is shown in 
Fig.~\ref{fig1}.  Our goal is to re-write these diagrams in such a way that all Feynman integrals 
can be dealt with using the integration-by-parts technique \cite{ibp}. To achieve this goal, 
we need to express the amplitude in terms of invariant form factors. To this end, we note that 
the most  general form of  ${\cal A}_{\mu \nu}^{(d)}$ is 
\be
{\cal A}_{\mu \nu}^{(d)}(p_1,p_3,p_4,p_2)  = \bar v_{p_2} \hat p_\perp u_{p_1} A^{(1,d)}_{\mu \nu} 
+ \bar v_{p_2} \gamma^\mu u_{p_1} A_\nu^{(2,d)} 
+ \bar v_{p_2} \gamma^\nu u_{p_1} A_\mu^{(3,d)} 
+ \bar v_{p_2} \gamma^{[\nu } \hat p_\perp \gamma^{\mu]} u_{p_1} A_4^{(4,d)},
\label{eq1.3}
\ee
where $ \gamma^{[\nu } \hat p_\perp \gamma^{\mu]} = \gamma^\nu  \hat p_\perp \gamma^\mu - \gamma^\mu  \hat p_\perp \gamma^\nu$
and the vector $p_\perp$ is defined by the Sudakov decomposition of the momenta $p_{3,4}$
\be
p_3 = \alpha_3 p_1 + \beta_3 p_2 + p_\perp,\;\;\;
p_4 = \alpha_4 p_1 + \beta_4 p_2 - p_\perp .
\label{sudec}
\ee
The transverse momentum $p_\perp$ is orthogonal to $p_{1,2}$,   $p_\perp \cdot p_{1,2} = 0$. 
The coefficients $\alpha_{3,4}$ and $\beta_{3,4}$ in Eq.(\ref{sudec}) can be 
written as 
\be
\alpha_3 = \frac{m_3^2 - u}{s},\;\;\; \beta_3 = \frac{m_3^2 - t}{s},\;\;\;
\alpha_4 = \frac{m_4^2 - t}{s},\;\;\; \beta_4 = \frac{m_4^2 - u}{s},
\ee
where we defined $p_3^2 = m_3^2$, $p_4^2 = m_4^2$ and introduced standard Mandelstam variables $ s=  (p_1 + p_2)^2,\;\;\
t = (p_1 - p_3)^2, \;\;\; u = (p_1 - p_4)^2$.  
The functions $A^{(i,d)},\;i = 1,...,4$, introduced in Eq.(\ref{eq1.3}),  depend on momenta and Lorentz indices. To make this dependence explicit, 
we decompose them into  invariant form factors $T_i,\; i=1,..,17$,
\be
\begin{split}
A_{\mu \nu}^{(1,d)} = &  T_1 g_{\mu \nu}  + T_2 p_{1\mu} p_{1\nu}
+T_3 p_{1\mu} p_{2\nu}+T_4 p_{1\mu} p_{\perp \nu}+T_5 p_{2\mu} p_{1\nu} 
+ T_6 p_{2\mu} p_{2\nu}+T_7 p_{2 \mu} p_{\perp \nu}
\\
& 
+T_8 p_{\perp \mu} p_{1 \nu} 
+T_9 p_{\perp \mu} p_{2 \nu}
+T_{10} p_{\perp \mu} p_{\perp \nu},\;\;\;\;
\\
A_{\mu}^{(2,d)} = &  T_{11} p_{1\mu} + T_{12} p_{2\mu} + T_{13} p_{\perp \mu},\;\;\;\;
A_{\nu}^{(3,d)} =  T_{14} p_{1\nu} + T_{15} p_{2\nu} + T_{16} p_{\perp \nu},\;\;\;
A^{(4,d)} = T_{17}.
\label{eq6}
\end{split}
\ee
We note that not all invariant form factors that appear in Eq.(\ref{eq6}) give 
independent contributions to physical amplitudes. This happens because  we 
did not use the transversality condition for lepton currents $p_3 \cdot \epsilon_3 = p_4 \cdot \epsilon_4 = 0$. 
As we will see shortly, when physical amplitudes are  computed, the number of relevant form factors 
will be reduced thanks to the transversality condition. 

To calculate the physical amplitude in Eq.(\ref{eq2}), 
we need  to contract ${\cal A}^{(d)}_{\mu \nu}$ with polarization vectors of external vector 
bosons.  As we already mentioned, they are given by matrix elements of the vector current  
between relevant leptonic states.  Using the spinor-helicity formalism, we write 
\be
\begin{split} 
& \epsilon_3^\mu(5_L) = \langle 5 |\gamma^\mu |6 ],\;\;\; \epsilon_3^\mu(5_R) = [ 5 |\gamma^\mu |6 \rangle 
= \langle 6 | \gamma^\mu | 5 ],
\\
& \epsilon_4^\nu(7_L) = \langle  7 |\gamma^\mu |8 ],\;\;\;\;
\epsilon_4^\nu(7_R) = [  7 |\gamma^\mu |8  \rangle = \langle 8 | \gamma^\mu | 7].
\label{eq7}
\end{split} 
\ee

Since, as we already pointed out, the helicity of the incoming quark fully determines the allowed 
helicity of the incoming antiquark, we need eight helicity amplitudes to fully describe 
production of two vector bosons. They are 
$
{\cal M}(q_L,5_L,7_L)$, ${\cal M}(q_L,5_R,7_L)$, ${\cal M}(q_L,5_L,7_R)$, ${\cal M}(q_L,5_R,7_R)$,
$ {\cal M}(q_R,5_L,7_L)$, ${\cal M}(q_R,5_R,7_L)$,
${\cal M}(q_R,5_L,7_R)$,  ${\cal M}(q_R,5_R,7_R)$.
We note that, according to Eq.(\ref{eq7}),  a change in  lepton helicities  can be obtained by 
interchanging momenta $p_5 \leftrightarrow p_6$ and $p_7 \leftrightarrow p_8$, where necessary.
Therefore,  we obtain  all required helicity amplitudes from ${\cal M}(q_L,5_L,7_L)$ 
and ${\cal M}(q_R,5_L,7_L)$ by  simple permutations of lepton momenta. 

For left- and right-handed  incoming quarks 
we find 
\be
\begin{split} 
& A_{LLL}^{(d)}(3,4) = {\cal A}_{\mu \nu}^{(d)}(p_1^{L},p_3,p_4,p_2^{R}) \epsilon_3^\mu (5_L) \epsilon_4^\nu(7_L)
=  
- F_1  \langle 5  7\rangle [ 8  6]    \langle 2 \hat 3 1] 
+  F_2 \langle  1  5\rangle  \langle 1  7\rangle [ 1  6 ][ 1  8]
  \langle 2 \hat 3 1]
\\
& \;\;\;\;\;\;
 + F_3 \langle 1  5\rangle \langle 2  7\rangle [ 16][ 2  8]  
 \langle 2 \hat 3 1]
+ F_5  \langle 1  7\rangle \langle 2  5\rangle [ 1  8] [ 2  6]  
   \langle 2 \hat 3 1]
 + F_6  \langle 2  5\rangle \langle 2  7\rangle [ 2  6] [ 2  8]  
  \langle 2 \hat 3 1] 
\\
 & \;\;\;\;\;\;
        + F_{11} \langle 2   5  \rangle \langle 1  7  \rangle  [ 1  6][ 1  8]  
+ F_{12} \langle 2  5\rangle  \langle 2  7\rangle  [ 1  6][ 2  8]  
+  F_{14} \langle 1  5\rangle  \langle 2  7\rangle  [ 1  6] [ 1  8]  
       + F_{15} \langle 2  5\rangle  \langle 2  7\rangle   [ 2  6]  [ 1  8 ],
\\
& A_{RLL}^{(d)}(3,4)  = {\cal A}_{\mu \nu}^{(d)}(p_1^{R},p_3,p_4,p_2^{L}) \epsilon_3^\mu (5_L) \epsilon_4^\nu(7_L)=
-F_1 \langle 57 \rangle [86] [2 \hat 3 1 \rangle 
 + F_2 \langle 1 5 \rangle \langle 1 7 \rangle [1 6] [1 8] [2 \hat 3 1 \rangle 
\\
&  \;\;\;\;\;\;
+ F_3 \langle 1  5\rangle \langle 2  7\rangle [ 16][ 2  8]  
 [ 2 \hat 3 1 \rangle 
+ F_5  \langle 1  7\rangle \langle 2  5\rangle [ 1  8] [ 2  6]  
   [ 2 \hat 3 1 \rangle 
 + F_6  \langle 2  5\rangle \langle 2  7\rangle [ 2  6] [ 2  8]  
  [ 2 \hat 3 1 \rangle  
\\
& \;\;\;\;\;\;
+F_{11}  [26] \langle 15 \rangle \langle 17 \rangle  [18] 
+ F_{12}   [28] \langle 15 \rangle [26] \langle 27 \rangle 
+ F_{14}  [28] \langle 15 \rangle [16] \langle 17 \rangle 
+F_{15} [28] \langle 25 \rangle [26]  \langle 17 \rangle ,
\label{eqampl}
\end{split} 
\ee
where the following {\it nine} combinations of form factors enter
\be
\begin{split} 
& F_1 = - 2T_1,\\
&F_2 =  T_2 - \alpha_3 \alpha_4 T_{10}- \alpha_3 T_8 + \alpha_4 T_4,  \\
& F_3 = T_3  - \frac{4 T_{17}}{s} - \alpha_3 \beta_4 T_{10} - \alpha_3 T_9 + \beta_4 T_4, \\
& F_5 = T_5 + \frac{4 T_{17}}{s} - \beta_3 \alpha_4 T_{10} - \beta_3 T_8 + \alpha_4 T_7,\\
& F_6 = T_6  - \beta_3 \beta_4 T_{10} - \beta_3 T_{9} + \beta_4 T_7,\\
& F_{11} = 2T_{11}  + 2\alpha_4T_{13},\\
& F_{12} = 2T_{12}  + 2\beta_4T_{13}, \\
& F_{14} = 2T_{14}  - 2 \alpha_3 T_{16},\\
& F_{15} = 2T_{15}  - 2 \beta_3 T_{16}.
\label{eq1.10}
\end{split} 
\ee
We note that form factors only depend on $s,t,u$ and $m_3^2,m_4^2$; therefore, they
are invariant under $p_5 \leftrightarrow p_6$ and $p_7 \leftrightarrow p_8$ 
permutations. For this reason,  it is straightforward to obtain all the relevant amplitudes from Eq.(\ref{eqampl}).
To find $T_{1,..,17}$ and $F_{1,2,..15}$, we 
need to construct projection operators.  To accomplish this, we write 
a generic amplitude as a matrix element of a string of Dirac matrices
\be
{\cal A}_{\mu \nu } = \bar v_{p_2} \hat \Gamma_{\mu\nu} u_{p_1}.
\ee
Multiplying it with $\bar u_{p_1} \hat O v_{p_2}$ and summing over spinor helicities, we find
\be
\sum {\cal A}_{\mu \nu } \bar u_{p_1} \hat O v_{p_2} = {\rm Tr} \left [ \hat p_2 \Gamma_{\mu \nu} \hat p_1 \hat O \right ].
\ee
Choosing different operators ${\cal O}$, we can project on individual $T$ form factors or their combinations. 
Below we  list all the projection operators and the results that we get when these operators  are convoluted with the amplitude
\be
\begin{split} 
& G_1 = - \frac{{\rm Tr} \left [ \hat p_2 \Gamma_{\mu \nu} \hat p_1 \hat p_\perp  \right ]}{4 p_\perp^2 ( p_1 \cdot p_2) ^3 } 
\times p_1^\mu p_1^\nu,\;\;\;\;\;\;\;G_1 = T_6,  \\
& G_2 = - \frac{{\rm Tr} \left [ \hat p_2 \Gamma_{\mu \nu} \hat p_1 \hat p_\perp  \right ]}{4 p_\perp^2 ( p_1 \cdot p_2) ^3 } 
\times p_2^\mu p_2^\nu,\;\;\;\;\;\;\;G_2 = T_2, \\
& G_3
= - \frac{{\rm Tr} \left [ \hat p_2 \Gamma_{\mu \nu} \hat p_1 \hat p_\perp  \right ]}{4 p_\perp^2 ( p_1 \cdot p_2) ^2 } 
\times p_1^\mu p_2^\nu,\;\;\;\;\;\;\;G_3 = T_1 + 2 T_{17} + (p_1 \cdot p_2) T_5  , \\
&  G_4 = - \frac{{\rm Tr} \left [ \hat p_2 \Gamma_{\mu \nu} \hat p_1 \hat p_\perp  \right ]}{4 p_\perp^2 ( p_1 \cdot p_2) ^2 } 
\times p_2^\mu p_1^\nu,\;\;\;\;\;\;\;G_4 = T_1 - 2 T_{17} + (p_1 \cdot p_2) T_{3}, \\
& G_5 = 
- \frac{{\rm Tr} \left [ \hat p_2 \Gamma_{\mu \nu} \hat p_1 \hat p_\perp  \right ]}{4 p_\perp^2 ( p_1 \cdot p_2) ^2 } 
\times p_1^\mu p_\perp^\nu,\;\;\;\;\;\;\;G_5 = T_7 p_\perp^2 + T_{15}, \\ 
& G_6 = 
- \frac{{\rm Tr} \left [ \hat p_2 \Gamma_{\mu \nu} \hat p_1 \hat p_\perp  \right ]}{4 p_\perp^2 ( p_1 \cdot p_2) ^2 } 
\times p_\perp^\mu p_1^\nu,\;\;\;\;\;\;\; G_6 = T_9 p_\perp^2 + T_{12}, \\ 
& G_7 = 
- \frac{{\rm Tr} \left [ \hat p_2 \Gamma_{\mu \nu} \hat p_1 \hat p_\perp  \right ]}{4 p_\perp^2 ( p_1 \cdot p_2) ^2 } 
\times p_\perp^\mu p_2^\nu,\;\;\;\;\;\;\; G_7 = T_8 p_\perp^2 + T_{11}, \\ 
& G_8 = 
- \frac{{\rm Tr} \left [ \hat p_2 \Gamma_{\mu \nu} \hat p_1 \hat p_\perp  \right ]}{4 p_\perp^2 ( p_1 \cdot p_2) ^2 } 
\times p_2^\mu p_\perp^\nu,\;\;\;\;\;\;\; G_8 = T_4 p_\perp^2 + T_{14}, \\ 
& G_9 = 
- \frac{{\rm Tr} \left [ \hat p_2 \Gamma_{\mu \nu} \hat p_1 \hat p_\perp  \right ]}{4 p_\perp^4 ( p_1 \cdot p_2) } 
\times p_\perp^\mu p_\perp^\nu,\;\;\;\;\;\;\; G_9 = T_1 + T_{10} p_\perp^2 + T_{13} + T_{16}, \\
& G_{10} = 
- \frac{{\rm Tr} 
\left [ \hat p_2 \Gamma_{\mu \nu} \hat p_1 \hat p_\perp  \right ]}{4 p_\perp^2 ( p_1 \cdot p_2)  } 
\times g^{\mu \nu},\;\;\;\;\;\;\;\;\; G_{10} = d \;T_1 + \left ( T_{3} + T_{5} \right ) p_1 \cdot p_2  + T_{10} p_\perp^2
+ T_{16} + T_{13},\\
& G_{11} = 
- \frac{{\rm Tr} \left [ \hat p_2 \Gamma_{\mu \nu} \hat p_1 \gamma^\mu  \right ]}{4 ( p_1 \cdot p_2) ^2 } 
\times p_2^\nu,\;\;\;\;\;\;\;\;\;\;\; G_{11} = T_8 \;p_\perp^2 + (d-2) T_{11},\\
& G_{12} = 
- \frac{{\rm Tr} \left [ \hat p_2 \Gamma_{\mu \nu} \hat p_1 \gamma^\mu  \right ]}{4 ( p_1 \cdot p_2) ^2 } 
\times p_1^\nu,\;\;\;\;\;\;\;\;\;\;\; G_{12} = T_9 \;p_\perp^2 + (d-2) T_{12},\\
& G_{13} = 
- \frac{{\rm Tr} \left [ \hat p_2 \Gamma_{\mu \nu} \hat p_1 \gamma^\nu  \right ]}{4 ( p_1 \cdot p_2) ^2 } 
\times p_1^\mu,\;\;\;\;\;\;\;\;\;\;\; G_{13} = T_7 \;p_\perp^2 + (d-2) T_{15}, \\
& G_{14} = 
- \frac{{\rm Tr} \left [ \hat p_2 \Gamma_{\mu \nu} \hat p_1 \gamma^\nu  \right ]}{4 ( p_1 \cdot p_2) ^2 } 
\times p_2^\mu,\;\;\; \;\;\;\;\;\;\;\;G_{14} = T_4 \;p_\perp^2 + (d-2) T_{14}, \\
& G_{15} = 
- \frac{{\rm Tr} \left [ \hat p_2 \Gamma_{\mu \nu} \hat p_1 \gamma^\mu  \right ]}{4 p_\perp^2 ( p_1 \cdot p_2)  } 
\times p_\perp^\nu,\;\;\; \;\;\;\;\;\;\;G_{15} = T_1 + T_{10} \;p_\perp^2 + (d-2) T_{13} + T_{16},
%\\
%\\
\end{split}
\ee
\be
\begin{split}
& G_{16} = 
- \frac{{\rm Tr} \left [ \hat p_2 \Gamma_{\mu \nu} \hat p_1 \gamma^\nu  \right ]}{4 p_\perp^2 ( p_1 \cdot p_2)  } 
\times p_\perp^\mu,\;\;\;\;\;\;\;\;\;\;\; G_{16} = T_1 + T_{10} \;p_\perp^2 + T_{13} + (d-2) T_{16},\\
& G_{17} = 
- \frac{{\rm Tr} \left [ \hat p_2 \Gamma_{\mu \nu} \hat p_1 
\gamma^{[\nu } \hat p_\perp \gamma^{\mu]}
%\left ( \gamma^\nu \hat p_\perp \gamma^\mu - \gamma^\mu \hat p_\perp \gamma^\nu \right )  
\right ]}{8 p_\perp^2 ( p_1 \cdot p_2)  },\;\;\;\;\;\;\;\;\; G_{17} = 
-2 (d - 2)(d-5)
%-(2 d^2-14d+20)
 T_{17} + (p_1 \cdot p_2) ( T_5 - T_3 ).
\nonumber 
\end{split}
\ee

Each of the $G_{1..17}$ projections  can be calculated from Feynman diagrams by taking traces and integrating 
over  loop momenta. These quantities depend on the Mandelstam invariants 
of the underlying $2 \to 2$ process $q  \bar q' \to V_1 V_2 $ but not on the polarization vectors
of electroweak  bosons  or on the momenta of final-state leptons. Thus, they can be  expressed in terms 
of Feynman integrals of the type introduced in Refs.~\cite{planar,nonplanar} and then reduced to master integrals 
using integration-by-parts identities. 

We can use projections $G_{1,..,17}$  to derive the $T$ form factors; the result reads 
\be
\label{eqTfromG}
\begin{split}
& T_1= \frac{G_{10}-G_{9}-G_{4}-G_{3}}{d-3},
\;\;\;\;\;\;\;\;\;\;T_2= G_2,
\\ 
& T_3 = -\frac{2( d(6-d)G_4 +(3-d)G_3
 +(d-4)G_{10}-7G_{4} +(4-d)G_{9}+G_{17})}{s_{12}(d-3) (d-4)},
\\
&T_4 = -\frac{G_{14}+(2-d)G_{8}}{p_\perp^2 (d-3)}, 
\\
&T_5 = -\frac{2((3-d)G_4-G_{17}   -(d^2 - 6d +7) G_3
+(d-4)G_{10}
 +(4-d)G_9)}{s_{12}(d-3)(d-4)},
\\
&T_6 = G_1,\;\;\;\;\;\;\;\;T_7 = -\frac{G_{13}+(2-d)G_{5}}{p_\perp^2 (d-3)},
\;\;\;\;\;\;\;
T_8 = - \frac{G_{11}+(2-d)G_7}{p_\perp^2 (d-3)},
\\
&T_9 = -\frac{G_{12}+(2 -d)G_{6}}{p_\perp^2 (d-3)},
\;\;\;\;\;\;\;\;
T_{10} = -\frac{G_{15}-G_{4}-G_{3}-d\;G_{9}+G_{16}+G_{10}}{p_\perp^2 (d-3)},
\\
&T_{11} = \frac{G_{11}-G_7}{d-3},\;\;\;\;\;\;
T_{12} = \frac{G_{12}-G_6}{d-3}, \;\;\;\;\;\;
T_{13} = \frac{G_{15}-G_9}{d-3},\;\;\;\; 
T_{14} = \frac{G_{14}-G_8}{d-3},
\\
&  T_{15} = \frac{G_{13}-G_5}{d-3},\;\;\;\;\; 
T_{16} = \frac{G_{16}-G_9}{d-3},\;\;\;\;\;\; 
T_{17} = -\frac{G_4-G_3+G_{17}}{2 (d-3)(d-4)}.
\end{split}
\ee
The $F$ form factors that are used for the evaluation of the 
amplitude can now be easily  computed using Eq.(\ref{eq1.10}).   We do not present these 
results here since they are not very illuminating. 

One point worth emphasizing though is that 
in Eq.(\ref{eqTfromG}) there are expressions  (see e.g. equation for $T_{17}$) that appear to have spurious  
singularities  in the limit $d \to 4$.  It would have been unfortunate if these 
singularities survive in the final formulas for form factors $F$ since, in such a situation, 
computation of all pieces needed for the evaluation of $G$'s, including master integrals,      to higher orders 
in $\epsilon$ is required.    It is therefore pleasing  
to observe that this does not happen and once results for the $F$ form factors are written in terms 
of  $G_{1,..,17}$ projectors, all spurious $(d-4)$ singularities disappear.  

For future reference, we give results for leading order form factors that appear in the physical amplitudes 
 Eq.(\ref{eqampl}) 
\be
\begin{split}
& F_{1} = \frac{2}{t},
\;\;\; F_2 = 0,
\;\;\; F_3 = -\frac{2}{st},
\;\; F_5 = \frac{2}{st},
\;\; F_6 = 0,
\;\;
F_{11} = \frac{2(m_{4}^2 - t)}{st},
\;\;\;
\\
& F_{12} = \frac{2( s +m_3^2 - t)}{st},
\;\;\;
F_{14} = -\frac{2( s +m_4^2 - t)}{st},\;\;\;
F_{15} = -\frac{2(m_3^2 -t )}{st}.
\end{split} 
\ee
We stress that these results are exact in a sense that no $d \to 4$ limit was taken to obtain them; in other words, 
all the $d$-dependence cancels out completely once physical form factors are computed. 

Before proceeding to  the discussion of various couplings of vector bosons to fermions  introduced in 
Eq.(\ref{eq2}), we want to explain why ${\cal A}^{(d)}$ and ${\cal A}^{(n_g)}$ amplitudes 
can be computed assuming that interactions of vector bosons with  quarks 
are mediated  by vector currents. The main reason is that 
we only consider here contributions of massless quarks and that, in such a case, the 
helicity is conserved. Therefore, we can always write couplings of vector bosons 
to fermions as linear combinations of left- and right-handed couplings and then  
move helicity projection  operators  to external lines. In this way the helicity-dependent 
couplings are generated and amplitudes that remain can be viewed as originating from  pure 
vector current interactions of gauge bosons  and quarks.

Of course, this discussion applies only to diagrams where vector bosons couple 
directly to  external  quark lines; in the notation of Eq.(\ref{eq2}) such diagrams 
contribute to ${\cal A}^{(d)}$ amplitudes.  However, we will now argue that {\it the same  
parametrization Eq.(\ref{eq1.3}) can be used to compute amplitudes ${\cal A}^{(n_g)}$}, 
which receive contributions from diagrams where vector bosons couple to closed fermion 
loops.  In fact,  this  would have been  obvious provided that electroweak  boson  
coupling to fermions is vector-like since in this case 
tracing over spin degrees of freedom in the internal quark loop gives us diagrams that are not very different from the ones 
that we already considered.  Potential problems could be expected 
 with axial couplings and, in particular, with terms 
where one axial and three vector couplings appear in $ggV_1 V_2$ Green's function. 
However, all such terms cancel because of $C$-parity conservation for massless 
fermions  for any  final state with two vector bosons \cite{ax}.\footnote{More precisely, for $C$-parity argument to be applicable, all fermions in the loop should have 
equal masses \cite{ax}.}  
Contributions of terms  with two vector and two axial couplings 
are not anomalous and must be equal to those  with four vector couplings. 
Hence, we conclude  that it is sufficient to consider  vector-current  
couplings of gauge  bosons to internal fermion loops,  to account for all non-vanishing contributions to the amplitude. 
The parametrization of the amplitude ${\cal A}^{(n_g)}$ is then taken from Eq.(\ref{eq1.3}) 
and its expression through  invariant form factors is taken from Eq.~(\ref{eqampl}).

To complete our construction of the scattering 
amplitude for $q \bar q' \to V_1 V_2$ in Eq.(\ref{eq2}), it remains to specify various 
helicity-dependent couplings that we introduced there. Below  we present these couplings 
for various pairs of electroweak gauge bosons that can be produced in proton collisions. 

\subsection{  $ \gamma^* \gamma^*$ production}

Photons are produced in the annihilation of a quark and an antiquark of the same flavor $q = q'$. 
Photon  interactions with both quarks and leptons are pure vector-like 
and, therefore, helicity-independent. We find 
\be
C^{L,R}_{q,\gamma} = -\sqrt{2} Q_q \sin \theta_W ,\;\;\;
C^{L,R}_{\bar q,\gamma} = -\sqrt{2} Q_q \sin \theta_W,
\;\;\;\;\;
\\
C^{L,R}_{l,\gamma}  = -\sqrt{2} Q_l \sin \theta_W,
\label{gamcoup}
\ee
where $Q_l$ and $Q_q$ are electromagnetic quark and  lepton charges 
in units of the positron charge. Finally, $C^{n_g}_{\gamma \gamma}$ is 
given by the sum of up and down quark charges of the first two generations
\be
C^{n_g}_{\gamma \gamma} = 2 \sin^2 \theta_W  \sum \limits_{i_q=1}^{2} 
\left ( Q_u^2 + Q_d^2 \right ) = \frac{20 \sin^2 \theta_W}{9}.
\ee

\subsection{ $ Z Z$ production}

Pairs of $Z$ bosons are produced in annihilation of quarks and anti-quarks of the same 
flavor. Couplings of $Z$ bosons to quarks and leptons
depend on  helicity and weak isospin of the corresponding particle. 
We find 
\be
C^{L,R}_{q ,Z} =  C^{L,R}_{\bar q ,Z} = 
\frac{1}{\sqrt{2} \cos \theta_W} \left ( V_q \pm  A_q \right ),\;\;\;
C^{L,R}_{l,Z} = \frac{1}{\sqrt{2} \cos \theta_W}  \left ( V_l \pm  A_l \right ),
\label{zcoup}
\ee
where 
\be
\begin{split} 
& V_u = \frac{1}{2} - \frac{4}{3} \sin^2 \theta_W,
\;\;\; A_u = \frac{1}{2},\;\;\;
V_d = -\frac{1}{2} + \frac{2}{3} \sin^2 \theta_W,
\;\;\; A_d = -\frac{1}{2},
\\
& V_e = -\frac{1}{2} + 2 \sin^2 \theta_W,
\;\;\; A_e = -\frac{1}{2},
\;\;\;\;
V_\nu = \frac{1}{2} 
\;\;\; A_\nu = \frac{1}{2}.
\label{eq18}
\end{split} 
\ee
The coefficient 
$C^{n_g}_{Z Z}$ is 
given by the sum of up and down quark vector and axial 
charges of the first two generations. We find 
\be
C^{n_g}_{Z Z} = \frac{1}{\cos^2 \theta_W}  \left ( V_u^2 + V_d^2 + A_u^2 + A_d^2 \right ). 
\ee

\subsection{ $ Z \gamma^*$ production}

Production of $Z$  and $\gamma^*$ occurs in annihilation of quarks and anti-quarks of the same 
flavor. The couplings to external fermions are given in Eqs.(\ref{gamcoup},\ref{zcoup}). The coefficient $C^{(n_g)}_{\gamma Z}$ 
reads
\be
C^{n_g}_{Z \gamma } = -\frac{2 \sin \theta_W}{\cos \theta_W }  \left ( V_u Q_u  + V_d Q_d \right ).
\ee

\subsection{ $W^+ \gamma^*$ and for $W^- \gamma^*$ production}

The $W^+ \gamma^*$ final state is produced in $u \bar d$ annihilation, $ u \bar d  \to W^+ \gamma^*$. 
Production of $W^- \gamma^*$ occurs in $d \bar u \to W^- \gamma^*$ process. 
Since $W$ bosons  interact with left-handed fermions, we have 
\be
\begin{split} 
& C^\lambda_{u,W^+} = \delta_{\lambda,L},\;\;\; 
C^{\lambda}_{\bar d,\gamma} = -\sqrt{2} Q_d \sin \theta_W,
\;\;\;
C^\lambda_{ u , \gamma} = -\sqrt{2} Q_u \sin \theta_W ,
\;\;\; C^\lambda_{\bar d, W^+} = \delta_{\lambda,L},\;\;\;
\\
& C^\lambda_{d,W^-} = \delta_{\lambda,L},\;\;\; 
C^{\lambda}_{\bar u,\gamma} = -\sqrt{2} Q_u \sin \theta_W,
\;\;\;
C^\lambda_{ d , \gamma} = -\sqrt{2} Q_d \sin \theta_W,
\;\;\; C^\lambda_{\bar u, W^-} = \delta_{\lambda,L}.
\\
& C^\lambda_{l,W^+} = C^\lambda_{l,W^-} = \delta_{\lambda,L},\;\;\;\;C^{\lambda}_{l,\gamma} = -\sqrt{2} Q_l \sin \theta_W.
\end{split} 
\ee
Coefficients  $C^{n_g}_{W^+\gamma}$ and $C^{n_g}_{W^- \gamma}$ vanish identically due to electric charge conservation.

\subsection{ $W^+ Z$ and for $W^-Z$ production}

Production processes for  $W^\pm Z$ are identical to $W^\pm \gamma^*$. The relevant couplings are
\be
\begin{split} 
& C^\lambda_{u,W^+} = \delta_{\lambda,L},\;\;\; C^\lambda_{\bar d, W^+} = \delta_{\lambda,L},\;\;\;,
C^\lambda_{d,W^-} = \delta_{\lambda,L},\;\;\; C^\lambda_{\bar u, W^-} = \delta_{\lambda,L},\;\;\; 
C^\lambda_{l,W^+} = C^\lambda_{l,W^-} = \delta_{\lambda,L},
\\
& C^{L,R}_{q ,Z} =  C^{L,R}_{\bar q ,Z} = 
\frac{1}{\sqrt{2} \cos \theta_W} \left ( V_q \pm  A_q \right ),\;\;\;
C^{L,R}_{l,Z} = \frac{1}{\sqrt{2} \cos \theta_W}  \left ( V_l \pm  A_l \right ),
\label{eq22}
\end{split} 
\ee
Coefficients 
$C^{n_g}_{W^\pm  Z}$ vanish identically due to  electric charge conservation.

\subsection{ $W^+W^-$ production}

Finally, we display the necessary couplings for the pair production of two $W$ bosons. 
This process can occur thanks to $u \bar u $ and $d \bar d$ annihilation.  
First, consider the process  $u(p_1) \bar u(p_2) \to W^+(p_3) W^-(p_4)$. 
In this case, 
\be
C_{u,W^+}^{\lambda} = \delta_{\lambda L},\;\;
 C_{\bar u, W^-}^{\lambda} = \delta_{\lambda,L},
\;\;\; C_{u,W^-}^{\lambda} = 0,
\;\;\;C_{\bar u,W^+}^{\lambda} = 0.
\ee
In case of $d(p_1) \bar d(p_2) \to W^+(p_3) W^-(p_4)$,
the coupling constants read 
\be
C_{d,W^+}^{\lambda} = 0,\;\;\;\;  
 C_{\bar d, W^-}^{\lambda} = 0, 
\;\;\; C_{d,W^-}^{\lambda} = \delta_{\lambda L},\;\;,
\;\;\;C_{\bar d,W^+}^{\lambda} = \delta_{\lambda,L}.
\ee
Couplings to leptons are given in Eq.(\ref{eq22}). 
The coefficient $C_{W^+W^-}^{(n_g)}$ receives contributions from two generations\footnote{As we already mentioned 
several times, we do not consider the third quark generation in this paper.} and reads 
\be
C_{W^+W^-}^{(n_g)} = 1. 
\ee

\section{Calculation of the form factors and the amplitudes} 
\label{results}

Once projection operators are  established, one can  compute the form factors and 
construct the scattering amplitudes for arbitrary di-boson final state. 
To this end, we note that all integrals that appear in the calculation of  form factors can be associated with 
one of the six topologies, introduced in Refs.~\cite{planar,nonplanar}.  Integrals that belong to each of these topologies 
are closed under the integration-by-parts identities \cite{ibp}. We use QGRAF \cite{qgraph} to generate Feynman diagrams 
and FORM \cite{form4} for algebraic manipulations and  computation of projections $G_{1,..,17}$.  We use FIRE 
\cite{Smirnov:2008iw,Smirnov:2013dia,Smirnov:fire}
to reduce all the integrals that appear 
in this calculation to master integrals. The master integrals were computed by us in Refs.~\cite{planar,nonplanar}.  
In principle, once all the ingredients are in place, computation of the amplitude becomes straightforward. In practice, however, 
it requires  some effort to put all the pieces together primarily because algebraic expressions that appear e.g. 
in the course of the reduction to master integrals are quite large  in size. The master integrals are expressed in terms 
of Goncharov polylogarithms up to weight four; we use GiNaC \cite{Bauer:2000cp} 
implementation  \cite{Vollinga:2004sn} to compute them. 

The results for the  amplitudes contain infra-red and ultraviolet divergences. The ultraviolet divergences 
are removed by the renormalization of the strong coupling constant. Since  tree-level scattering 
amplitudes are independent of $\alpha_s$, we only need its renormalization through one loop. It 
 reads 
\be
\alpha_s^{(0)} S_\ep = \alpha_s \mu^{2\ep} 
\left ( 1- \frac{\beta_0}{\ep} \left ( \frac{\alpha_s}{2 \pi} \right ) 
+ {\cal O}(\alpha_s^2) \right ),
\ee
where $\alpha_s^{(0)}$ is the bare and $\alpha_s = \alpha_s(\mu)$ is the renormalized coupling constant. 
In addition, $\beta_0 = (11 C_A -4 T_R n_f)/6 = 11/2 - n_f/3$ is the QCD beta-function, 
$n_f$ is the number of massless 
flavors, $C_A = 3, T_R = 1/2$ and  
$S_\ep = (4\pi)^\ep e^{-\ep \gamma_E}$, with  $\gamma_E$ being  the Euler constant.

According to Catani \cite{Catani:1998bh}, infra-red singularities of UV-renormalized amplitudes at next-to-next-to-leading order 
are fully determined by 
leading and next-to-leading order amplitudes. This feature can be used as an important check of the 
correctness of the computation.  To introduce Catani's result, we write  a UV-renormalized 
amplitude as 
\be
{\cal M} = {\cal M}_{0} + \left ( \frac{\alpha_s(\mu)}{ 2\pi } \right ) {\cal M}_1 
+ \left ( \frac{\alpha_s(\mu)}{2 \pi}  \right )^2 {\cal M}_2 +....
\ee
Leading order amplitude ${\cal M}_0$ in this formula is finite and, in fact, $d$-independent. 
The next-to-leading order amplitude contains infra-red divergences that, however, can be written 
in a factorized form 
\be
{\cal M}_1 = \hat I_1(\ep) {\cal M}_0 + {\cal M}_{\rm fin}.
\ee
For the process of vector boson pair production, the final state is neutral. Therefore,  
\begin{equation}
I_1(\epsilon) = - \frac{e^{-\epsilon(L_s-\gamma_E)}}{\Gamma(1-\epsilon)} C_F\left(\frac{1}{\epsilon^2}+\frac{3}{2 \epsilon}\right),
\end{equation}
where $L_S =  \ln (-s/\mu^2 -i0) = \ln|s/\mu^2| - i\pi$ and $C_F=4/3$.  In variance with ${\cal M}_0$, ${\cal M}_1$ and 
${\cal M}_{1,\rm fin}$ depend on the dimensional regularization parameter $\epsilon$. This feature is 
important for proper comparison of Catani's formula with results of explicit computation.

\begin{table}[t]
\begin{center}
\begin{tabular}{|l|c|c|c|c|c|}
\hline
Helicity& $\; \ep^{-2} \;$ & $\ep^{-1}$ & $\ep^0$ & $\ep$ & $\ep^2$ \\
\hline
$ {\tilde A}^{(d,1)}_{LLL} $ &  
$-\frac{4}{3} $ & $-2 - i4.1887902$ &  
$ 3.2003253 - i\;6.2069828  $ & 
$ 5.4520124 - i\;2.9495550 $ & 
$ 5.3607865 - i\;3.1814370 $ 
\\
\hline
$ {\tilde A}^{(d,1)}_{RLL} 
$ &  $-\frac{4}{3} $ & 
$-2 -i4.1887902$ & 
$  4.0059079  - i\; 3.0593301 $ & 
$  0.5752861  + i\; 5.8440713 $ & 
$ -9.6769949  + i\; 0.9775875 $ 
\\
\hline
\end{tabular}
\end{center}
\caption{
Ratios of selected  one-loop  helicity amplitudes and  tree amplitudes, see Eq.(\ref{eq34}).
Momenta of external particles 
are given in the main text of the paper. 
Tree-amplitudes are ${\cal A}^{(d,0)}_{LLL}=  4600.82746 - i\; 17933.17244 $
and ${\cal A}^{(d,0)}_{RLL}=  732.100366 - i\; 1148.55597 $.  
 }
\label{table:oneloop}
\end{table}

The two-loop amplitude ${\cal M}_2$ can be written in a similar way 
\be
{\cal M}_2 = \hat I_2(\ep) {\cal M}_0 + \hat I_1(\ep)  {\cal M}_1 + {\cal M}_{2,\rm fin},
\ee
where 
\be
\hat I_2 = -\frac{1}{2} I_1^2(\ep) - \frac{\beta_0}{\ep} I_1(\ep) 
	    + \frac{e^{-\ep \gamma_E} \Gamma(1-2\ep)}{\Gamma(1-\ep)} \left ( \frac{\beta_0}{\ep} + k_q \right ) I_1(2\ep) +
	    \frac{H_q}{2\ep}.
\ee
The two constants that enter this formula read 
\be
\begin{split} 
& k_q = \left ( \frac{67}{18} - \frac{\pi^2}{6} \right ) C_A - \frac{10}{9} T_R n_f, \\
& 
H_q = C_F^2 \left (-\frac{3}{8} + \frac{\pi^2}{2} - 6\zeta_3 \right ) + 
 C_F n_f T_R \left (-\frac{25}{54} + \frac{\pi^2}{12} \right ) 
    +  C_F C_A \left ( \frac{245}{216} - \frac{23 \pi^2}{48} + \frac{13 \zeta_3}{2} \right ).
\end{split} 
\ee

We note that  Catani formula can  also be used for individual  form factors that we introduced earlier to describe physical amplitudes.
 To this end, we only need  to replace tree and loop amplitudes in the above formulas 
with the corresponding form factors.  We have used the above results for the infra-red poles of scattering amplitudes 
to check the correctness of our computation of the amplitudes ${\cal A}^{(d)}$.  We also note that the amplitude 
${\cal A}^{(n_g)}$ appears for the first time at NNLO; therefore, according to Catani's formula it cannot have 
infra-red $1/\ep$ singularities.  This is an important check of the correctness of the computation of the amplitude $A^{(n_g)}$.

\begin{table}[t]
\begin{center}
\begin{tabular}{|l|c|c|c|c|c|}
\hline
Helicity& $\; \ep^{-4} \;$ & $\ep^{-3}$ & $\ep^{-2}$ & $\ep^{-1}$ & $\ep^0$ \\
\hline
$ {\tilde A}^{(d,2)}_{LLL}$ &  
$  \frac{8}{9} $ & 
$  1.388889 + i\;  5.58505$ & 
$ -16.02478 + i\;  8.625043 $ & 
$ -29.43232 - i\; 28.086442 $ & 
$  23.53917 - i\; 67.620386 $ \\ 
\hline
$ {\tilde A}^{(d,2)}_{RLL} $ &  
$ \frac{8}{9}  $ & 
$   1.388889 + i\;   5.58505 $ & 
$  -17.09889 + i\;   4.42818 $  & 
$   -8.268265- i\;  37.414997 $ & 
$   73.483267 + i\; 23.301609 $ 
 \\
\hline
\end{tabular}
\end{center}
\caption{
Ratios of selected two-loop  helicity amplitudes and  tree amplitudes, see Eq.(\ref{eq34}).
Momenta of external particles 
are given in the main text of the paper. The number of massless fermion species $n_f$ is taken to be five.
Tree-amplitudes are ${\cal A}^{(d,0)}_{LLL}=  4600.82746 - i\; 17933.17244 $
and ${\cal A}^{(d,0)}_{RLL}=  732.100366 - i \; 1148.55597 $.  
 }
\label{table:twoloop}
\end{table}

We turn to the discussion of  numerical results for the scattering amplitudes ${\cal A}^{(d)}$, ${\cal A}^{(n_g)}$.
We define those amplitudes by contracting them with polarization vectors of electroweak bosons
\be
{\cal A}^{(d,n_g)}(\lambda_q,\lambda_5,\lambda_7) = 
{\cal A}^{(d,n_g)}(p_1^{\lambda_q},p_3^{\lambda_5},p_4^{\lambda_7},p_2^{-\lambda_q}) 
= {\cal A}_{\mu \nu}^{(d,n_g)}(p_1^{\lambda_q},p_3,p_4,p_2^{-\lambda_q})  \epsilon^\mu_3(\lambda_5) \epsilon^\nu_4(\lambda_7). 
\ee
and write their perturbative expansion as 
\be
\begin{split} 
& {\cal A}^{(d)} = 
 {\cal A}^{(d,0)}
\left [1  + a_0 \; s^{-\ep} \tilde{ A}^{(d,1)}
+ 
a_0^2\; s^{-2 \ep} \tilde{ A}^{(d,2)}
\right ],\;\;\;\;\; 
{\cal A}^{(n_g)} = 
 {\cal A}^{(d,0)} 
a_0^2 \; s^{-2 \ep} \tilde{ A}^{(n_g,2)},
\end{split}  
\label{eq34}
\ee
where $a_0 = \alpha_s^{(0)}  (4\pi)^{\ep} \Gamma(1+\ep)/(2\pi)$.
We note that in Eq.(\ref{eq34}) we choose to expand in bare, rather than renormalized, QCD coupling. Also, we made it 
explicit in Eq.(\ref{eq34}) that 
the amplitude ${\cal A}^{(n_g)}$ appears  at two loops for the first time.

To motivate our choice of  kinematics for numerical results for  
the amplitudes that we present below, we  consider  $q \bar q' \to W^+W^-$ production as the background to Higgs boson 
signal  in $pp \to H \to W^+W^-$. Therefore,  we choose the center-of-mass energy $\sqrt{s}$ to be  the 
mass of the Higgs boson $\sqrt{s} = m_H = 125~{\rm GeV}$. The invariant mass of the vector boson $V_1$ is set to 
$p_3^2 = m_W^2$, with $m_W =  80.419~{\rm GeV}$. The invariant mass of the second vector boson $V_2$ is set to $25~{\rm GeV}$.
We take the vector boson scattering angle in the center-of-mass collision frame to be $\pi/3$ radians. 
We also take decay angles of the lepton $l_5$ in the rest frame of the boson $V_1$ to be $\theta_5 = \pi/4$ 
and $\varphi_5 = \pi/2$ and decay angles of the lepton $l_7$ in the rest frame of the boson $V_2$ 
to be $\theta_7 = \pi/6$ and $\varphi_7 = \pi$.  The four-momenta of initial and final state 
particles are given by 
\be
\begin{split}
& p_1 = (  62.5,0,0,62.5),\;\;\;\;\;  p_2 = (62.5,0,0,-62.5),\\
& p_5 = ( 
   48.2561024468725,\;        13.8697156788798,\;       -28.4324101181205,\;     
   36.4400941989053  ), \\ 
& p_6 = ( 
   37.6127597971275,\;        12.2010429705974,\;        28.4324101181205,\;     
  -21.3881346746519     
), \\     
& p_7 = ( 
   19.5655688780000 ,\;      -19.2853793247386,\;     0,\;
   3.29933778517879     
),\\     
& p_8 = (
  19.5655688780000,\;       -6.78537932473856, \;      0, \;
  -18.3512973094322 
).  
\end{split} 
\ee
To obtain numerical results for the amplitude, we take the number of massless fermion species $n_f$ to be five. 
Results for selected  helicity amplitudes $A^{(d)}$ and $A^{(n_g)}$ are 
shown in Tables~\ref{table:oneloop},\ref{table:twoloop},\ref{table:twoloopng}. They  can be compared to predictions 
based on Catani's formula\footnote{Catani's formula needs to be re-written to provide an expansion 
in the unrenormalized QCD coupling.}. For both one- and two-loop amplitudes,  divergent terms agree perfectly. 
For the one-loop amplitudes, we can also compare the ${\cal O}(\ep^0)$ terms against 
the known results \cite{oneloopampl}; we find perfect agreement. The amplitude $A^{(n_g)}$ that describes contributions 
to $q \bar q' \to V_1V_2$ where vector bosons couple to closed loops of fermions  
is finite as expected
 since those amplitudes have no tree- and one-loop contributions.

\begin{table}[t]
\begin{center}
\begin{tabular}{|l|c|c|c|c|c|}
\hline
Helicity& $\; 1/\ep^4 \;$ & $1/\ep^3$ & $\ep^2$ & $1/\ep$ & $\ep^0$ \\
\hline
${\tilde A}^{(n_g,2)}_{LLL} $ &  
$0$ &
${\cal O}(10^{-14})$ & 
${\cal O}(10^{-10})$  & 
${\cal O}(10^{-8})$ &
$ -0.4339650 - i\; 0.16264039 $
 \\
\hline
$ {\tilde A}^{(n_g,2)}_{RLL} $ &  
$0$ &
${\cal O}(10^{-13})$ & 
${\cal O}(10^{-8})$  & 
${\cal O}(10^{-7})$ &
$ 4.3238033 + i \; 1.8252651$
 \\
\hline
\end{tabular}
\end{center}
\caption{
Ratios of finite two-loop  ${\cal A}^{(n_g)}$ helicity amplitudes to tree amplitudes. 
Momenta of external particles 
are given in the main text of the paper. 
Tree-amplitudes are ${\cal A}^{(d,0)}_{LLL}=  4600.82746 - i\; 17933.17244 $
and ${\cal A}^{(d,0)}_{RLL}=  732.100366 - i\; 1148.55597 $.  
 }
\label{table:twoloopng}
\end{table}

We will now elaborate on the numerical stability of our results. This is an important issue 
since amplitudes for 
 vector boson pair productions may exhibit numerical instabilities in the limit of forward or 
backward scattering. In fact, numerical stability depends on the vector boson  transverse momentum 
since $1/p_\perp^2$  singularities appear when Feynman integrals are reduced to master integrals. 
We have  evidence from previous studies about values of transverse momenta where  such instabilities  arise. 
In  case of {\it one-loop} $gg \to VV$ amplitudes, numerical instabilities 
start to appear for transverse momenta of an order of a few {\rm GeV} \cite{Kauer:2012hd,Campbell:2013una} 
and sophisticated treatment is required to remove them completely \cite{Campbell:2013una}.  To explore 
numerical (in)stability of our results, we study amplitudes 
${\tilde A}^{(d,n_g,2)}_{LLL} $  in dependence of the vector boson scattering angle.  All other kinematic 
variables  are taken to be identical to what we described above.

\begin{figure}[t]
  \centering
  \includegraphics[angle=-90,width=0.45\textwidth]{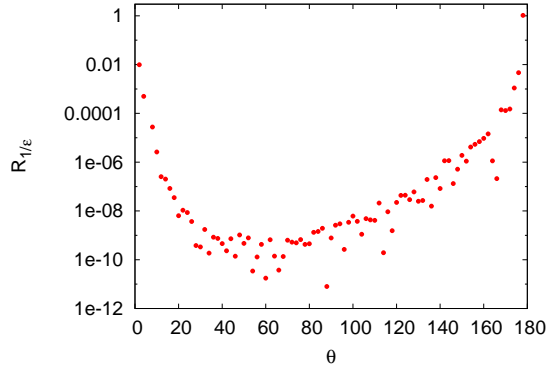}
  \caption{
Absolute value squared of the ratio of $1/\ep$ poles of the $LLL$ scattering amplitude computed from 
Feynman diagrams and using Catani formula,
$R_{1/\ep} = | |{\tilde A}^{(d,2)}_{LLL}|^2/|{\tilde A}^{(d,2)}_{C,LLL}|^2 - 1|$. 
 The center-of-mass energy, 
gauge boson masses and lepton scattering angles are given in the  text. }
  \label{fig2}
\end{figure}

In  Fig.~\ref{fig2} we show ratios of $1/\ep$ singularities in the squared amplitude $|{\tilde A}^{(d,2)}_{LLL}|^2$
computed directly and using Catani's formula. Deviations of this ratio from one signal numerical instabilities. 
We observe that $|{\tilde A}^{(d,2)}_{LLL}|^2/|{\tilde A}^{(d,2)}_{C,LLL}|^2 - 1$ is of order 
${\cal O}(10^{-6}-10^{-10})$ in the bulk of the phase-space. Significant instabilities are observed 
for backward scattering ($178$ degrees), where the transverse momentum is close to $1~{\rm GeV}$. 
However, the situation improves considerably already for $176$ degree scattering where the transverse 
momentum is $2~{\rm GeV}$.  The forward scattering limit appears to be more stable; even at two  degrees, 
the $1/\ep$ contribution is computed properly to within a percent. 

\begin{figure}[t]
  \centering
  \includegraphics[angle=-90,width=0.45\textwidth]{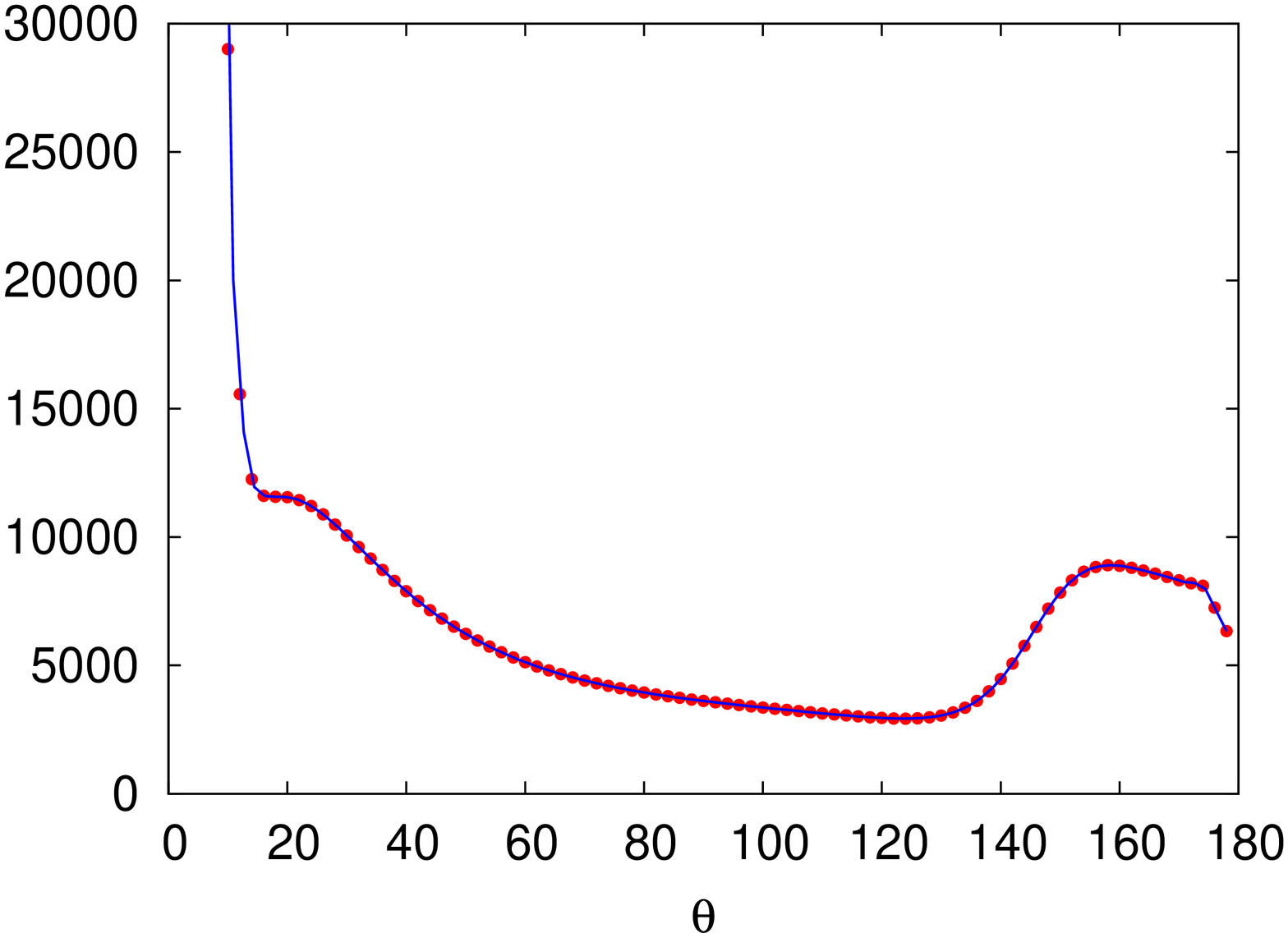}\;\;\;\;;
  \includegraphics[angle=-90,width=0.45\textwidth]{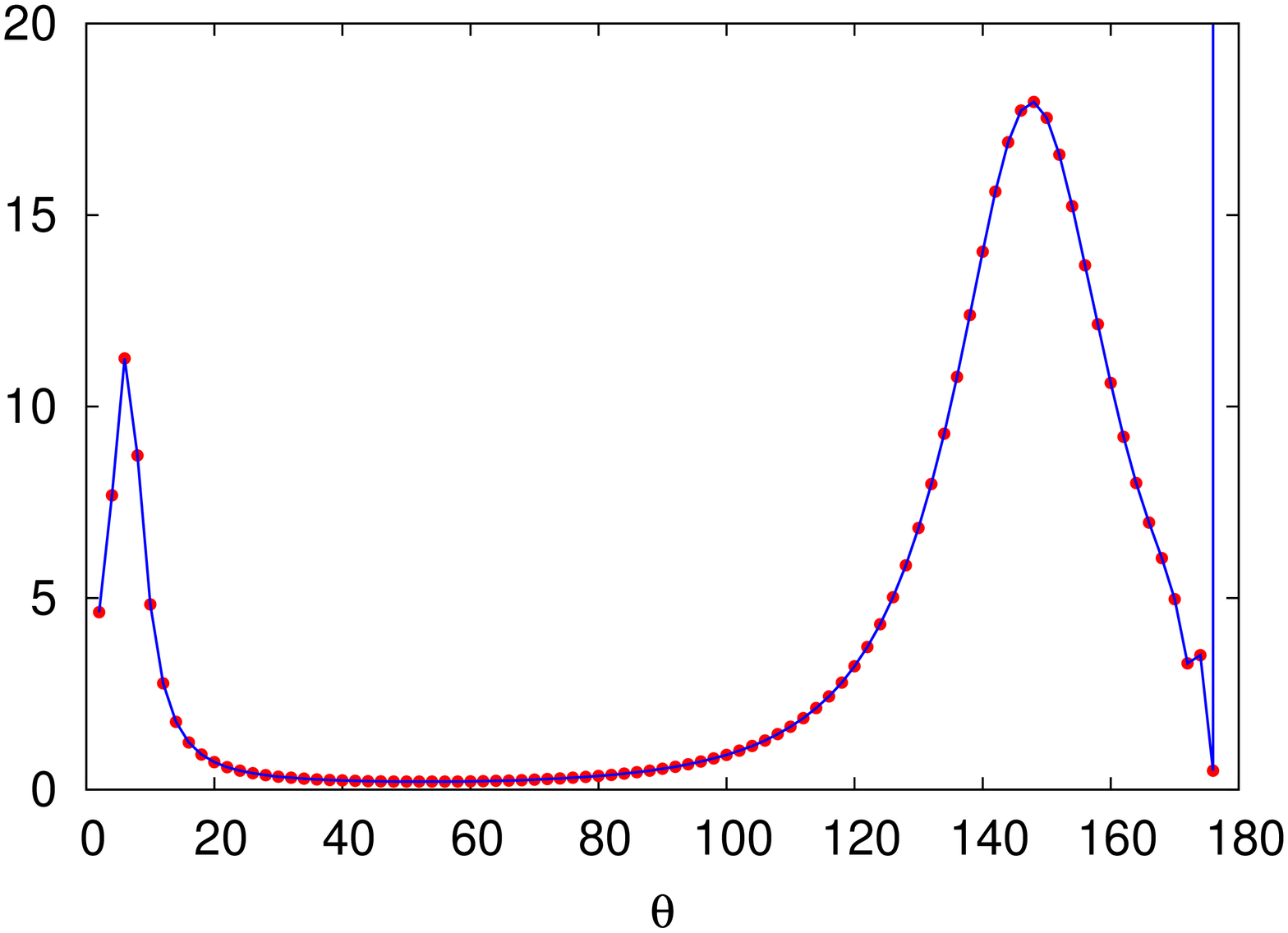}
  \caption{
 Absolute values squared of  
${\tilde A}^{(d,2)}_{LLL}$ ( left pane) and 
${\tilde A}^{(n_g,2)}_{LLL}$ (right pane) 
as a function   of the vector boson scattering angle. The center-of-mass energy, 
gauge boson masses and lepton scattering angles are given in the  text. }
  \label{fig3}
\end{figure}

In the left  pane of Fig.~\ref{fig3} we show the  absolute value squared of the ratio of the finite part of the ${\cal A}^{(d,2)}_{LLL}$ amplitude 
and the leading order amplitude ${\cal A}^{(d,0)}_{LLL}$.   To understand  numerical accuracy 
of these results, we compared the output obtained with the double-precision version of the Fortran code
with the Mathematica implementation. 
The advantage of the latter is that it provides a possibility to compute amplitudes with  
arbitrary numerical  precision thereby ameliorating the problem of numerical instability. 
For backward scattering, we find that up to $174$ degrees, the finite part is computed to within a per mille.
For forward direction, the situation is similar but, perhaps, slightly better. 

Finally, in the right pane of Fig.~\ref{fig3} we show 
 absolute value squared of the ratio of the finite part of the left amplitude $A^{(n_g,2)}_{LLL}$ 
 and the leading order $LLL$ amplitude ${\cal A}^{(d,0)}_{LLL}$. 
Numerical  instabilities are apparent for backward scattering. In fact, at $170$ degrees ($p_\perp \sim 5~{\rm GeV}$), the agreement between 
double-precision Fortran code and the Mathematica code is about ten percent.  In the forward direction,
the situation is much better --  the double-precision Fortran results agree with the results obtained using 
 Mathematica implementation  to better than a fraction of a percent for scattering angles as small as six degrees.

To summarize, while the two-loop amplitudes that we compute in this paper do exhibit  numerical instabilities 
at small values of vector boson transverse momenta, we believe the stability is acceptable for phenomenological 
applications. Moreover, there are several ways to improve the situation. For example, it is possible to 
extend the Fortran code to provide results with quadruple precision. Note that computation of master 
integrals with arbitrary precision is feasible since Goncharov polylogarithms implementation in GiNaC 
does provide this functionality \cite{Vollinga:2004sn}.  
Moreover, it should  also be possible to construct  expansion of analytic expressions 
for  scattering  amplitudes that we obtained in this paper 
around singular limits, for example for forward or backward scattering and  threshold production. If such expansions become available, 
computation of helicity amplitudes in singular limits will  be significantly simplified.  We leave these improvements for  future work.

\section{Conclusions} 
\label{concl}

In this paper we described  computation of two-loop scattering amplitudes for the annihilation of a 
quark and an antiquark  into four leptons,  that occurs through the production of two  electroweak gauge bosons. 
The invariant masses of gauge bosons are kept arbitrary. 
We have given  explicit formulas  for  projection operators that allow  one to compute contributions of individual 
Feynman diagrams to invariant form factors.  We use 
these form factors to construct helicity amplitudes for vector boson pair production processes including all off-shell 
effects and leptonic decays of vector bosons.

Results for two-loop  scattering amplitudes 
obtained in this paper remove the last obstacle for computing the NNLO QCD corrections 
to the production of pairs of vector bosons with identical and different invariant masses.  The two-loop virtual 
corrections that we compute in this paper will have to be combined  with one-loop amplitudes
for $pp \to V_1 V_2 j$ and with tree amplitudes for  $pp \to V_1 V_2 jj$.  While doing 
this consistently is non-trivial, the relevant technology is well-understood by now
\cite{nnlo_tech}. We hope, therefore, that 
results for NNLO fiducial volume 
cross sections for pair production of electroweak bosons 
can be expected in the near future. 

\section{Acknowledgments} 
J.M.H. and A.V.S. are grateful to Institute for Theoretical Particle Physics (TTP) at  Karlsruhe Institute of Technology 
 for the hospitality extended to them 
during the work on this paper. 
Research of K.M and F.C. was  partially supported by US NSF under grant PHY-1214000.
K.M. was also  partially  supported by Karlsruhe Institute of Technology through its distinguished 
researcher fellowship program. J.M.H. is supported in part by the DOE grant DE-SC0009988 and by 
the Marvin L. Goldberger fund.  F.C. and A.S. are supported in part by DFG through SFB/TR 9.
V.S. is  supported in part by the Alexander von Humboldt Foundation (Humboldt Forschungspreis).

{\bf Note added} In the original version of this paper, one of the form factors that contributes to the amplitude ${\cal A}^{n_g}$ 
was calculated incorrectly. We are grateful L.~Tancredi for pointing this out to us. After correcting this mistake, 
our results for the two-loop amplitudes $q \bar q \to V_1 V_2$ agree with the results of an  independent calculation in Ref.~\cite{tancredi}.


\begin{thebibliography}{99}

\bibitem{atlas7} G. Aad {\it et al.} ATLAS collaboration, Phys. Rev. D{\bf 87}, (2013), 112001 [Erratum 
ibid., {\bf 88}, (2013) 079906].

\bibitem{cms7} CMS Collaboration, CMS-PAS-SMP-12-005.

\bibitem{cms8} CMS Collaboration, CMS-PAS-SMP-12-013. 

\bibitem{bsm} 
D.~Curtin, P.~Meade, P.J. Tien, arXiv:1406.0848;
D. Curtin, P. Jaiswal, P. Meade,  Phys.\ Rev.\ D {\bf 87}  (2013)  031701.


%\cite{Meade:2014fca}
\bibitem{Meade:2014fca} 
  P.~Meade, H.~Ramani and M.~Zeng,
  %``Transverse momentum resummation effects in W^+W^- measurements,''
  arXiv:1407.4481 [hep-ph].
  %%CITATION = ARXIV:1407.4481;%%
  %1 citations counted in INSPIRE as of 17 Aug 2014


%\cite{Jaiswal:2014yba}
\bibitem{Jaiswal:2014yba} 
  P.~Jaiswal and T.~Okui,
  %``An Explanation of the WW Excess at the LHC by Jet-Veto Resummation,''
  arXiv:1407.4537 [hep-ph].
  %%CITATION = ARXIV:1407.4537;%%
  %1 citations counted in INSPIRE as of 17 Aug 2014

\bibitem{limits} For recent results and earlier references, see e.g. 
CMS Collaboration, hep-ex/1406.0113; CMS collaboration Phys. Rev. {\bf D89} (2014) 092005,
ATLAS collaboration, Phys. Lett. B {\bf 717}  (2012)  49.

\bibitem{Cascioli:2014yka} 
  F.~Cascioli, T.~Gehrmann, M.~Grazzini, S.~Kallweit, P.~Maierh\"ofer, A.~von Manteuffel, S.~Pozzorini, 
D.~Rathlev, L. Tancredi and  E. Weihs,
  %``ZZ production at hadron colliders in NNLO QCD,''
  arXiv:1405.2219 [hep-ph].

\bibitem{babis} 
Ch. Anastasiou, J. Cancino, F. Chavez, C. Duhr, A. Lazopoulos, B. Mistlberger and  R. Mueller, 
 arXiv:1408.4546 [hep-ph].

\bibitem{tg1} 
   T.~Gehrmann, M.~Grazzini, S.~Kallweit, P.~Maierh\"ofer, A.~von Manteuffel, S.~Pozzorini, 
D.~Rathlev, L. Tancredi,
  arXiv:1408.5243 [hep-ph].

\bibitem{nloqcd} J. Ohnemus and J. F. Owens, 
Phys. Rev. D {\bf 43} (1991) 3626;
 J. Ohnemus, Phys. Rev. D {\bf 44} (1991) 3477;
J. Ohnemus,  Phys. Rev. D
{\bf 44} (1991) 1403;
B. Mele, P. Nason and G. Ridolfi,  Nucl. Phys. B {\bf 357} (1991) 409;
S. Frixione, P. Nason and G. Ridolfi, Nucl. Phys. B {\bf 383} (1992) 3;
S. Frixione,  Nucl. Phys. B {\bf 410} (1993) 280;
U. Baur, T. Han and J. Ohnemus, Phys. Rev.  D {\bf 53} (1996)
1098;
L. J. Dixon, Z. Kunszt and A. Signer, Phys. Rev. D
{\bf 60}, 114037 (1999);
J. M. Campbell and R. K. Ellis,  Phys. Rev. D {\bf 60} (1999) 113006;
J. M. Campbell, R. K. Ellis and C. Williams,  JHEP {\bf 1107}, 018 (2011).

\bibitem{shower} S. Frixione and B. R. Webber, JHEP {\bf 0206} (2002) 029; 
P. Nason and G. Ridolfi, JHEP {\bf 0608} (2006) 077;
K. Hamilton,  JHEP 1101 (2011) 009;
T. Melia, P. Nason, R. Rontsch and G. Zanderighi, JHEP {\bf 1111} (2011) 078;
R. Frederix, S. Frixione, V. Hirschi, F. Maltoni, R. Pittau and P. Torrielli,
 JHEP {\bf 1202} (2012) 099;
  F.~Cascioli, S.~Höche, F.~Krauss, P.~Maierhöfer, S.~Pozzorini and F.~Siegert,
  %``Precise Higgs-background predictions: merging NLO QCD and squared quark-loop corrections to four-lepton + 0,1 jet production,''
  JHEP {\bf 1401}  (2014) 046;
  F.~Campanario, M.~Rauch and S.~Sapeta,
  %``$W^+W^-$ production at high transverse momenta beyond NLO,''
  Nucl.\ Phys.\ B {\bf 879} (2014) 65. 


\bibitem{nloew} M. Billoni, S. Dittmaier, B. J\"ager and C. Speckner, 
JHEP {\bf 1312} (2013) 043; A. Bierweiler, T. Kasprzik, H. K\"uhn and S. Uccirati, JHEP {\bf 1211} (2012) 093;
A. Bierweiler, T. Kasprzik and J. H. K\"uhn, JHEP {\bf 1312} (2013) 071;
J. Baglio, L. D. Ninh and M. M. Weber,  Phys. Rev. D {\bf 88} (2013) 113005.

\bibitem{resum}   M. Grazzini,  JHEP {\bf 0601} (2006) 095;
S. Dawson, I. M. Lewis and M. Zeng,  Phys.
Rev. D {\bf 88} (2013) 5, 054028;
 Y. Wang, C. S. Li, Z. L. Liu, D. Y. Shao and H. T. Li, Phys. Rev.
D {\bf 88} (2013) 114017.






\bibitem{ax}  E.W.N.~Glover and J.J. van der Bij, Phys. Lett.  B {\bf 219}, (1989) 488;
G.~Kao and D.A~Dicus, Phys. Rev. D {\bf 43}, (1991), 1555. 



\bibitem{gglo}
T. Binoth, M. Ciccolini, N. Kauer and M. Kr\"amer, , JHEP {\bf 0503} (2005) 065;
T. Binoth, M. Ciccolini, N. Kauer and M. Kr\"amer,  JHEP {\bf 0612} (2006) 046.


\bibitem{nnlo_tech}
  A.~Gehrmann-De Ridder, T.~Gehrmann and E.~W.~N.~Glover,  JHEP {\bf 0509}, 056 (2005);
  S.~Catani and M.~Grazzini,  Phys.\ Rev.\ Lett.\  {\bf 98}, 222002 (2007);
  S.~Catani, L.~Cieri, D.~de Florian, G.~Ferrera and M.~Grazzini,   Nucl.\ Phys.\ B {\bf 881}, 414 (2014);
  S.~Weinzierl,  JHEP {\bf 0303}, 062 (2003);
  G.~Somogyi, P.~Bolzoni and Z.~Trocsanyi,  Nucl.\ Phys.\ Proc.\ Suppl.\  {\bf 205-206}, 42 (2010) and references therein.
  M.~Czakon,
  %``A novel subtraction scheme for double-real radiation at NNLO,''
  Phys.\ Lett.\  B {\bf 693}, 259-268 (2010);
%  [arXiv:1005.0274 [hep-ph]].
  M.~Czakon,
  %``Double-real radiation in hadronic top quark pair production as a proof of a certain concept,''
  Nucl.\ Phys.\  B {\bf 849}, 250-295 (2011);
  M.~Czakon and D.~Heymes,
  %``Four-dimensional formulation of the sector-improved residue subtraction scheme,''
  arXiv:1408.2500 [hep-ph];
  R.~Boughezal, K.~Melnikov and F.~Petriello,
  %``A subtraction scheme for NNLO computations,''
  Phys.\ Rev.\ D {\bf 85}, 034025 (2012).


\bibitem{Gehrmann:2014bfa} 
  T.~Gehrmann, A.~von Manteuffel, L.~Tancredi and E.~Weihs,
  %``The two-loop master integrals for $q\overline{q} \to VV$,''
  JHEP {\bf 1406}, 032 (2014).
  %%CITATION = ARXIV:1404.4853;%%
  %7 citations counted in INSPIRE as of 19 Aug 2014



\bibitem{Gehrmann:2013cxs} 
  T.~Gehrmann, L.~Tancredi and E.~Weihs,
  %``Two-loop master integrals for $q \bar{q} \to VV$: the planar topologies,''
  JHEP {\bf 1308},  (2013) 070;
  %%CITATION = ARXIV:1306.6344,;%%
  %20 citations counted in INSPIRE as of 19 Aug 2014

\bibitem{nonplanar} 
F.~Caola, J. M. Henn, K.~Melnikov, V. A. Smirnov, arXiv:1404.5590.


\bibitem{planar} 
 J.~M.~Henn, K.~Melnikov and V.~A.~Smirnov,
  %``Two-loop planar master integrals for the production of off-shell vector bosons in hadron collisions,''
  JHEP {\bf 1405} (2014) 090.



\bibitem{ffnnlo} 
R.J. Gonsalves, Phys. Rev. D {\bf 28} (1983) 1542;
W.L. van Neerven, Nucl. Phys. B {\bf 268} (1986) 453;
G. Kramer and B. Lampe, J. Math. Phys. 28 (1987) 945.


\bibitem{ibp}  F. Tkachov, Phys. Lett. B {\bf 100}  (1981)  65; 
K.G.~Chetyrkin and F.~Tkachov, Nucl. Phys. B {\bf 192} (1981)  159.



\bibitem{jhenn} 
J.M.~Henn, Phys. Rev. Lett. {\bf 110} (2013), 251601.


\bibitem{qgraph} P. Nogueira, J.Comput.Phys. 105 (1993) 279.

\bibitem{form4} J. Kuipers, T. Ueda, J. Vermaseren, and J. Vollinga, 
Comput.Phys.Commun. 184 (2013) 1453.

\bibitem{Smirnov:2008iw}
  A.~V.~Smirnov,
  %``Algorithm FIRE -- Feynman Integral REduction,''
  JHEP {\bf 0810} (2008) 107.
  %%CITATION = ARXIV:0807.3243;%%

\bibitem{Smirnov:2013dia}
  A.~V.~Smirnov and V.~A.~Smirnov,
  %``FIRE4, LiteRed and accompanying tools to solve integration by parts relations,''
  Comput.\ Phys.\ Commun.\  184 (2013) 2820.
  %%CITATION = ARXIV:1302.5885;%%


\bibitem{Smirnov:fire}
  A.~V.~Smirnov, arXiv:1408.2372 [hep-ph].


%\cite{Bauer:2000cp}
\bibitem{Bauer:2000cp}
  C.~W.~Bauer, A.~Frink and R.~Kreckel,   cs/0004015 [cs-sc].
  %%CITATION = CS/0004015;%%
  %75 citations counted in INSPIRE as of 20 Apr 2014


%\cite{Vollinga:2004sn}
\bibitem{Vollinga:2004sn}
  J.~Vollinga and S.~Weinzierl,
  %``Numerical evaluation of multiple polylogarithms,''
  Comput.\ Phys.\ Commun.\  167   (2005) 177.
  %%CITATION = HEP-PH/0410259;%%
  %82 citations counted in INSPIRE as of 20 Apr 2014


%\cite{Catani:1998bh}
\bibitem{Catani:1998bh} 
  S.~Catani,
  %``The Singular behavior of QCD amplitudes at two loop order,''
  Phys.\ Lett.\ B {\bf 427} (1998) 161.
%  [hep-ph/9802439].
  %%CITATION = HEP-PH/9802439;%%
  %307 citations counted in INSPIRE as of 18 Aug 2014



\bibitem{oneloopampl}
L. J. Dixon, Z. Kunszt and A. Signer,  Nucl. Phys. B {\bf 531}
(1998) 3;

%\cite{Kauer:2012hd}
\bibitem{Kauer:2012hd} 
  N.~Kauer and G.~Passarino,
  %``Inadequacy of zero-width approximation for a light Higgs boson signal,''
  JHEP {\bf 1208} (2012) 116.
  %%CITATION = ARXIV:1206.4803;%%
  %73 citations counted in INSPIRE as of 22 Aug 2014


%\cite{Campbell:2013una}
\bibitem{Campbell:2013una} 
  J.~M.~Campbell, R.~K.~Ellis and C.~Williams,
  %``Bounding the Higgs width at the LHC using full analytic results for $gg -> e^- e^+ \mu^- \mu^+$,''
  JHEP {\bf 1404} (2014) 060.
  %%CITATION = ARXIV:1311.3589;%%
  %25 citations counted in INSPIRE as of 22 Aug 2014


\bibitem{tancredi} 
T.~Gehrmann, A.~von Manteuffel,  L. Tancredi, arXiv:1503.04812.

\end{thebibliography}
\end{document}